\documentclass[10pt,conference]{IEEEtran}

\makeatletter
\def\ps@headings{%
\def\@oddhead{\mbox{}\scriptsize\rightmark \hfil \thepage}%
\def\@evenhead{\scriptsize\thepage \hfil \leftmark\mbox{}}%
\def\@oddfoot{}%
\def\@evenfoot{}}
\def\blfootnote{\xdef\@thefnmark{}\@footnotetext} 
\makeatother
\usepackage{graphicx}
\usepackage{dcolumn}
\usepackage{bm}
\usepackage{epsfig}
\usepackage{psfrag}
\usepackage{subfigure}
\usepackage{color}
\usepackage{amsmath, amssymb, latexsym,amsfonts,epsfig, graphicx, verbatim,stmaryrd}
\usepackage[nospace]{cite}
\usepackage{algorithmic}
\usepackage{algorithm}
\usepackage{multirow}

\newtheorem{lemma}{Lemma}

\newcommand{\eqn}[1]{Eq.(#1)}
\newcommand{\av}[1]{\langle #1 \rangle}

\newcommand{\eq}{\!\!=\!}
\newcommand{\m}{\!-\!}
\newcommand{\Prob}{\mathbb{P}}
\newcommand{\Mean}{\mathbb{E}}

\newcommand{\est}[1]{\widehat{#1}}
\newcommand{\RW}{{\scriptscriptstyle\textrm{RW}}}
\newcommand{\MHRW}{{\scriptscriptstyle\textrm{MH}}}
\newcommand{\BFS}{{\scriptscriptstyle\textrm{BFS}}}
\newcommand{\RMSE}{RMSE}
\newcommand{\real}{{\scriptscriptstyle\textrm{real}}}
\newcommand{\AV}{{\scriptstyle\textrm{av}}}
\newcommand{\TOT}{{\scriptstyle\textrm{tot}}}

\newcommand{\eg}{\emph{e.g.}, }
\newcommand{\ie}{\emph{i.e.}, }

\begin{document}

\title{Towards Unbiased BFS Sampling}


\author{\IEEEauthorblockN{Maciej Kurant}
\IEEEauthorblockA{EECS Dept\\
University of California, Irvine\\
{\em maciej.kurant@gmail.com}}
\and
\IEEEauthorblockN{Athina Markopoulou}
\IEEEauthorblockA{EECS Dept\\
University of California, Irvine\\
{\em athina@uci.edu}}
\and
\IEEEauthorblockN{Patrick Thiran}
\IEEEauthorblockA{School of Computer \& Comm. Sciences\\
EPFL, Lausanne, Switzerland\\
{\em patrick.thiran@epfl.ch}}
}

\maketitle

\begin{abstract}
Breadth First Search (BFS) is a widely used approach for sampling large unknown Internet topologies.
Its main advantage over random walks and other exploration techniques 
is that a BFS sample is a plausible graph on its own, 
and therefore we can 
study its topological characteristics. 
However, it has been empirically observed that incomplete BFS is biased toward high-degree nodes, which may strongly affect the measurements. 


In this paper, we first analytically quantify the degree bias of BFS sampling. In particular, we calculate the node degree distribution expected to be observed by BFS as a function of the fraction~$f$ of covered nodes, in a random graph $RG(p_k)$ with an arbitrary degree distribution~$p_k$.
We also show that, for $RG(p_k)$, all commonly used graph traversal techniques (BFS, DFS, Forest Fire, Snowball Sampling, RDS) suffer from exactly the same bias.

Next, based on our theoretical analysis, we propose a practical BFS-bias correction procedure. 
It takes as input a collected BFS sample together with its fraction~$f$. 
Even though $RG(p_k)$ does not capture many graph properties common in real-life graphs (such as assortativity), 
our $RG(p_k)$-based correction technique performs well on a broad range of Internet topologies and on two large BFS samples of Facebook and Orkut networks. 

Finally, we consider and evaluate a family of alternative 
correction procedures, and demonstrate that, although they are unbiased 
for an arbitrary topology, their large variance makes them far less effective than the $RG(p_k)$-based technique. 
\end{abstract}

\section{Introduction}

\blfootnote{This paper is a revised and extended version of~\cite{Kurant2010}. 
}

A large body of work in the networking community focuses on Internet topology measurements at various levels, including the IP or AS connectivity, the Web (WWW), peer-to-peer (P2P) and online social networks (OSN).
The size of these networks and other restrictions make measuring the entire graph impossible. 
For example, learning only the topology of Facebook social graph would require downloading more than $250TB$ of HTML data~\cite{Gjoka2010,Gjoka2011_Facebook_JSAC}, which is most likely impractical.
Instead, researchers typically collect and study a small but representative sample of the underlying graph.

In this paper, we are particularly interested in sampling networks that naturally allow to explore the neighbors of a given node (which is the case in WWW, P2P and OSN). A number of graph exploration techniques
use this basic operation for sampling. They can be roughly classified in two categories: 
(i)~random walks, and (ii)~graph traversals.

In the first category, \emph{random walks}, nodes can be revisited. This category includes the classic Random Walk (RW)~\cite{Lovasz93} and its variations~\cite{Ribeiro2010,Avrachenkov2010}, as well as the Metropolis-Hastings Random Walk (MHRW). They are used for sampling of nodes on the Web~\cite{Henzinger2000}, P2P networks~\cite{Stutzbach2006-unbiased-p2p,Gkantsidis2004,Rasti09-RDS}, OSNs~\cite{Gjoka2010, Twitter08} and large graphs in general~\cite{Leskovec2006_sampling_from_large_graphs}.
Random walks are well studied~\cite{Lovasz93} and result in samples that have either no bias (MHRW) or a known bias (RW) that can be corrected for~\cite{Feld1991,Newman01_EgoCentered_Networks,Salganik2004,VolzHeckathorn08}.
In contrast to BFS, random walks collect a representative sample of nodes rather than of topology, and are therefore {\em not the focus of the paper}. However, we use them as baseline for comparison.

\begin{figure}[t]
\psfrag{y0}[r][c][0.9]{$\av{q_k}$ }
\psfrag{y1}[l][c][0.8]{expected observed}
\psfrag{y2}[l][c][0.8]{average node degree}
\psfrag{k3}[r][c][0.9]{$\langle k\rangle$}
\psfrag{k2}[r][c][0.9]{$\frac{\langle k^2\rangle}{\langle k\rangle}$}
\psfrag{i}[c][c][0.9]{$f$\ \ fraction of sampled nodes}
\psfrag{tot}[c][c][0.9]{1}
\psfrag{Zero}[c][c][0.9]{0}
\psfrag{rw}[c][c][0.7]{Random Walk (RW)}
\psfrag{a}[l][c][0.7]{Graph traversal techniques:}
\psfrag{b}[l][c][0.7]{- BFS}
\psfrag{c}[l][c][0.7]{- DFS}
\psfrag{d}[l][c][0.7]{- Forest Fire}
\psfrag{e}[l][c][0.7]{- Snowball / RDS}
\psfrag{mhrw}[c][c][0.7]{Metropolis-Hastings Random Walk (MHRW)}
\includegraphics[width=0.49\textwidth]{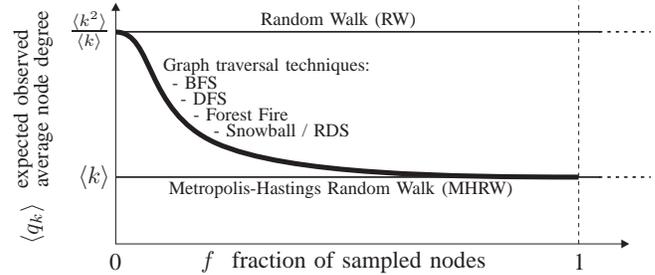}
\caption{{\bf Overview of analytical results.} We calculate the node degree distribution~$q_k$ expected to be observed by BFS in a random graph $RG(p_k)$ with a given degree distribution~$p_k$, as a function of the fraction of sampled nodes~$f$. 
(In this plot, we show only its average~$\av{q_k}$.)  We show RW and MHRW as a reference.
$\langle k\rangle = \av{p_k}$ is the real average node degree, and $\langle k^2\rangle$ is the real average squared node degree.
\quad~{\em Observations:}
\quad~(1)~For a small sample size, BFS has the same bias as RW; with increasing $f$, the bias decreases; a complete BFS ($f\eq 1$) is unbiased,  as is MHRW (or uniform sampling).
\quad~(2)
All common graph traversal techniques (that do not revisit the same node) lead to the same bias.
\quad~(3)~The shape of the BFS curve depends on the real node degree distribution~$p_k$, but it is always monotonically decreasing; we calculate it precisely in this paper.
\quad~(4)~We also calculate the original distribution $p_k$ based on the sampled~$q_k$ and $f$ (not shown here). 
}
\label{fig:contributions}
\end{figure}

In the second category, \emph{graph traversals}, each node is visited exactly once (if we let the process run until completion and if the graph is connected). These methods vary in the order in which they visit the nodes; examples include BFS, Depth-First Search (DFS), Forest Fire (FF), Snowball Sampling (SBS) and Respondent-Driven Sampling (RDS)\footnote{RDS is essentially SBS equipped with some bias correction procedure (omitted in Fig.~\ref{fig:contributions}).}. 
Graph traversals, especially BFS, are very popular and widely used for sampling Internet topologies, \eg in WWW~\cite{Najork01} or OSNs~\cite{Ahn-WWW-07, Mislove2007, Wilson09}. 
\cite{Mislove2007} alone has about 380 citations as of December 2010, many of which use its Orkut BFS sample. 
%
The main reason of this high popularity is that 
a BFS sample is a plausible graph on its own. 
Consequently, we can study its topological characteristics (\eg shortest path lengths, clustering coefficients, community structure), 
which is a big advantage of BFS over random walks.
%
Of course, this approach is correct only if the BFS sample is representative of the entire graph. At first sight it seems true, \eg a BFS sample of a lattice is a (smaller) lattice.

Unfortunately, this intuition often fails. It was observed empirically that BFS introduces a bias towards high-degree nodes~\cite{Najork01,Lee-Phys-Rev-06,snowball-bias,Ye2010}. We also confirmed this fact in a recent measurement of Facebook~\cite{Gjoka2010,Gjoka2011_Facebook_JSAC}, where our BFS crawler found the average node degree $324$, while the real value is only~$94$. 
This means that 
the average node degree is overestimated by BFS by about 250\%! This has a striking effect not only on the node property statistics, but also on the topological metrics.


Despite the popularity of BFS on the one hand, and its bias on the other hand, we still know relatively little about the statistical properties of node sequences returned by BFS. The formal analysis is challenging because BFS, similarly to every sampling without replacement, introduces complex dependencies between the sampled nodes difficult to deal with mathematically.

\smallskip
\emph{Contributions. } 
Our work is a step towards understanding the statistical characteristics of BFS samples and correcting for their biases, with the following main contributions.

First, we focus on a random graph $RG(p_k)$ with a given (and arbitrary) degree distribution $p_k$. We calculate precisely the node degree distribution~$q_k$ expected to be observed by BFS as a function of the fraction~$f$ of sampled nodes. 
We illustrate this and related results in Fig.~\ref{fig:contributions}.
To the best of our knowledge, this is the first analytical result describing the bias of BFS sampling. 

Second, based on our theoretical analysis, we propose a practical BFS-bias correction procedure. 
It takes as input a collected BFS sample together with the fraction~$f$ of covered nodes, and estimates the mean of an arbitrary function $x(v)$ defined on graph nodes. 
Even though $RG(p_k)$ misses many graph properties common in real-life graphs (such as assortativity), 
our $RG(p_k)$-based correction technique performs well on a broad range of Internet topologies, and on two large BFS samples of Facebook and Orkut networks. 
We make its ready-to-use \texttt{python} implementation publicly available at~\cite{kurant_networkx_traversals}. 

Third, we complement the above findings by proposing a family of alternative correction procedures that are unbiased for any arbitrary topology. Although seemingly attractive, they are characterized by large variance, which makes them far less effective than the $RG(p_k)$-based correction technique.



\smallskip
{\em Scope. }
Our theoretical results hold strictly for the random graph model~$RG(p_k)$. (However, we show that they apply relatively well to a broad range of real-life topologies.) 
We also restrict our attention to static graphs with self-declared unweighted social links; 
dynamically varying graphs~\cite{Stutzbach2006-unbiased-p2p, Stutzbach2006, Rasti2008, MisloveWosn08, Latapy2008a, Rasti09-RDS, Willinger09-OSN_Research,Magnien2009} 
and interaction graphs~\cite{Valafar2009a, Schneider2009a,Viswanath2009}
are out of the scope of this paper.

Finally, our $RG(p_k)$-based bias-correction procedure is designed for local graph properties, such as node statistics. 
Our analytical results can potentially help the estimation of non-local graph properties (such as graph diameter), which is our main direction for the future.

\smallskip
{\em Outline. }
The outline of the paper is as follows. Section~\ref{sec:Related_Work} discusses related work. 
Section~\ref{sec:Algorithms} presents BFS and other graph traversal algorithms under study. We also briefly describe random walks that are used as baseline for comparison throughout the paper. 
Section~\ref{sec:Graph model} presents the random graph~$RG(p_k)$ model used in this paper. 
Section~\ref{sec:Analysis} analyzes the degree bias of BFS. 
Section~\ref{sec:Correcting for node degree bias} shows how to correct for this bias.
Section~\ref{sec:Simulation} evaluates our results in simulations and by sampling real world networks. 
Section~\ref{sec:Unbiased BFS estimators} introduces and evaluates alternative BFS-bias correction techniques.
Section~\ref{sec:Recommendations} gives some practical sampling recommendations, and Section~\ref{sec:Conclusion} concludes the paper. 

\section{Related Work}\label{sec:Related_Work}

{\em BFS used in practice.} BFS is widely used today for exploring large networks, such as OSNs. 
In \cite{Ahn-WWW-07}, Ahn et al.   used BFS to sample Orkut and MySpace. In \cite{Mislove2007} and \cite{MisloveWosn08}, Mislove et al.  used BFS  to crawl the social graph in four popular OSNs: Flickr, LiveJournal, Orkut, and YouTube. 
\cite{Mislove2007} alone has about 380 citations as of December 2010, many of which use its highly biased Orkut BFS sample.
In \cite{Wilson09}, Wilson et al. measured the social graph and the user interaction graph of Facebook using several BFSs, each BFS constrained in one of the largest 22 regional Facebook networks. In our recent work~\cite{Gjoka2010,Gjoka2011_Facebook_JSAC}, we have also crawled Facebook using various sampling techniques, including BFS, RW and MHRW.

{\em BFS bias.}
It has been empirically observed that incomplete BFS and its variants introduce bias towards high-degree nodes~\cite{Najork01}\cite{Lee-Phys-Rev-06,snowball-bias,Ye2010}. 
We confirmed this in Facebook~\cite{Gjoka2010,Gjoka2011_Facebook_JSAC}, which, in fact, inspired and motivated this paper.
Analogous bias has been observed in the field of social science, for sampling techniques closely related to BFS, \ie Snowball Sampling and RDS~\cite{Goodman61_Snowball_sampling,Heckathorn97_RDS_introduction,Salganik2004} (see Section~\ref{subsec:SBS}).

{\em Analyzing BFS.} 
To the best of our knowledge, the sampling bias of BFS has not been analyzed so far.
\cite{Kim06_poisson_cloning} and \cite{Achlioptas05_On_the_bias_of_traceroute_sampling} are the closest related papers to our methodology. The original paper by Kim~\cite{Kim06_poisson_cloning} analyzes the size of the largest connected component in classic Erd\"os-R\'enyi random graph by essentially applying the configuration model with node degrees chosen from a Poisson distribution. 
To match the stubs (or ``clones'' in~\cite{Kim06_poisson_cloning}) uniformly at random in a tractable way, Kim proposes a ``cut-off line'' algorithm.  He first assigns each stub a random index from $[0,np]$, and next progressively scans this interval. 
Achlioptas et al. used this powerful idea in~\cite{Achlioptas05_On_the_bias_of_traceroute_sampling} to study the bias of traceroute sampling in random graphs with a given degree distribution.
The basic operation in \cite{Achlioptas05_On_the_bias_of_traceroute_sampling} is traceroute (\ie ``discover a path'') and is performed from a single node to all other nodes in the graph.
The union of the observed paths forms a ``BFS-tree'', which includes all nodes but misses some edges (\eg those between nodes at the same depth in the tree). In contrast, the basic operation in the traversal methods presented in our paper is to discover all neighbors of a node, and it is applied to all nodes in increasing distance from the origin. 
Another important difference
is that~\cite{Achlioptas05_On_the_bias_of_traceroute_sampling} studies a completed BFS-tree, whereas we study the sampling process when it has visited only a fraction $f<1$ of nodes. Indeed, a completed BFS ($f\eq 1$) is trivial in our case: it has no bias, as all nodes are covered.

In the field of social science, a significant effort was put to correct for the bias of BFS's close cousin - Snowball Sampling (SBS)~\cite{Goodman61_Snowball_sampling}. SBS together with a bias correction procedure is called Respondent-Driven Sampling (RDS)~\cite{Heckathorn97_RDS_introduction}. The currently used correction technique~\cite{Salganik2004,VolzHeckathorn08} assumes that nodes can be revisited, which essentially approximates SBS by Random Walk (see Section~\ref{subsec:Random Walk (RW)}). 
In this paper, we formally show that this approximation is valid if the fraction~$f$ of sampled nodes is relatively small. However, as~\cite{Gile2009} points out, the current RDS methodology is systematically biased for larger~$f$. Consequently, \cite{Gile2011} proposed an SBS bias correction method based on the random graph $RG(p_k)$. This is essentially the same basic starting idea as used in our original paper published independently~\cite{Kurant2010}. However, the two papers fundamentally differ in the final solution: \cite{Gile2011} proposes a simulation-aided approach, whereas we solve the problem analytically.

Another recent and related paper is~\cite{Illenberger09_snowball_bias_correction}. 
The authors propose and evaluate a heuristic approach to correct the degree bias in the $i$th generation of SBS, based on the values measured in the generation~$i\m 1$. 
In practice, this generation-based scheme may be challenging to implement, because the number of nodes per generation may grow close to exponential with~$i$. Consequently, we are likely to face a situation where collecting the next generation is prohibitively expensive, while the current generation has much fewer nodes than our sampling capabilities allow for.

\smallskip
{\em Probability Proportional to Size Without Replacement (PPSWOR).}
At a closer look, our $RG(p_k)$-based approach reduces BFS (and other graph traversals) to a classic sampling design called Probability Proportional to Size Without Replacement (PPSWOR)~\cite{Yates1953,Raj1956,Murthy1957,Hartley1962,Andreatta1986,Rao1991,Kochar2001,Fattorini2006}.
Unfortunately, to the best of our knowledge, none of the existing results is directly applicable to our problem.
This is because, speaking in the terms used later in this paper, the available results either (i)~require the knowledge of $q_k(f)$ (expected, not sampled) as an input,  (ii)~propose how to calculate $q_k(f)$ for the first two nodes only, or (iii)~calculate $q_k(f)$ as an average of many simulated traversals of the known graph (in contrast, we only have one run on unknown graph)~\cite{Fattorini2006}.
In fact, this work can be naturally extended to address the problems with PPSWOR.

\smallskip
{\em Previous version of this paper.}
This work is a revised and extended version of our recent conference paper~\cite{Kurant2010}. 
The main changes are: 
(i)~a successful application of our $RG(p_k)$-based correction procedure to a wide range of large-scale real-life Internet topologies (Table~\ref{tab:Real-life topologies}, Fig.~\ref{fig:known_topologies}, Fig.~\ref{fig:unknown_topologies}(d), Section~\ref{subsec:Real-life fully known topologies}),
(ii)~bias correction procedures for arbitrary node properties (Section~\ref{sec:Correcting for node degree bias}), 
(iii)~a complementary BFS-bias correction technique (Section~\ref{sec:Unbiased BFS estimators}), and 
(iv)~a publicly available ready-to-use \texttt{python} implementation of our approach. 

Finally, we would like to stress that our two other JSAC submissions~\cite{Gjoka2011_Facebook_JSAC,Gjoka2011_multigraph_JSAC} focus on sampling techniques based on random walks, which differ in fundamental aspects (sampling with replacement vs without, sampling of nodes vs of topology) from the BFS sampling addressed here.

\section{Graph exploration techniques}\label{sec:Algorithms}

Let $G=(V,E)$ be a connected graph with the set of vertices~$V$, and a set of undirected edges $E$.
Initially, $G$ is unknown, except for one (or some limited number of) seed node(s).
When sampling through graph exploration, we begin at the seed node, and we recursively visit (one, some or all) its neighbors.
We distinguish two main categories of exploration techniques: random walks and graph traversals. 

\subsection{Random walks (baseline)}\label{subsec:Exploration with replacements}
Random walks allow revisiting the same node many times. We consider\footnote{We include random walks only as a useful baseline for comparison with graph traversals (\eg BFS). The analysis of random walks does not count as a contribution of this paper.} the following classic examples:

\smallskip
\subsubsection{Random Walk (RW)}
In this classic sampling technique~\cite{Lovasz93}, we start at some seed node. At every iteration, the next-hop node~$v$ is chosen uniformly at random among the neighbors of the current node~$u$. It is easy to see that RW introduces a linear bias towards nodes of high degree~\cite{Lovasz93}.

\smallskip
\subsubsection{Metropolis Hastings Random Walk (MHRW)}

In this technique, as in RW, the next-hop node~$w$ is chosen uniformly at random among the neighbors of the current node~$u$. However, with a probability that depends on the degrees of $w$ and~$u$, MHRW performs a self-loop instead of moving to $w$.
More specifically, the probability $P^\MHRW_{u,w}$ of moving from $u$ to $w$ is as follows~\cite{mcmc-book}:
\begin{equation}\label{eq:P_u,w}
    P^\MHRW_{u,w} = \left\{ \begin{array}{ll}
\frac{1}{k_u} \cdot \min(1, \frac{k_u}{k_w}) & \textrm{if $w$ is a neighbor of $u$,} \\
1- \sum_{y\neq u} P^\MHRW_{u,y} & \textrm{if $w=u$,} \\
0 & \textrm{otherwise},
\end{array} \right.
\end{equation}
where $k_v$ is the degree of node $v$. Essentially, MHRW reduces the transitions to high-degree nodes and thus eliminates the degree bias of RW. This property of MHRW was recently exploited in various network sampling contexts \cite{Stutzbach2006-unbiased-p2p,Twitter08,Gjoka2010,Rasti09-RDS}.

\subsection{Graph traversals}
In contrast, 
graph traversals never revisits the same node. At the end of the process, and assuming that the graph is connected, all nodes are visited. 
However, when using graph traversals for sampling, we terminate after having collected a fraction~$f<1$ (usually $f\ll 1$) of graph nodes.  

\smallskip
\subsubsection{Breadth First Search (BFS)}
BFS is a classic graph traversal algorithm that starts from the seed and progressively explores all neighbors. 
At each new iteration the earliest explored but not-yet-visited node is selected next. Consequently, BFS discovers first the nodes closest to the seed.

\smallskip
\subsubsection{Depth First Search (DFS)}
This technique is similar to BFS, except that at each iteration we select the latest explored but not-yet-visited node. As a result, DFS explores first the nodes that are faraway (in the number of hops) from the seed.

\smallskip
\subsubsection{Forest Fire (FF)}\label{sec:Forest Fire (FF)}
FF is a randomized version of BFS, where for every neighbor $v$ of the current node, we flip a coin, with probability of success $p$, to decide if we explore $v$. FF reduces to BFS for $p\eq 1$. It is possible that this process dies out before it covers all nodes. In this case, in  order to make FF comparable with other techniques, we revive the process from a random node already in the sample. Forest Fire is inspired by the graph growing model of the same name proposed in~\cite{Leskovec05_Forest_Fire} and is used as a graph sampling technique in~\cite{Leskovec2006_sampling_from_large_graphs}.

\smallskip
\subsubsection{Snowball Sampling (SBS) and Respondent-Driven Sampling (RDS)}\label{subsec:SBS}

According to a classic definition  by Goodman~\cite{Goodman61_Snowball_sampling}, an $n$-name Snowball Sampling is  similar to BFS, but at every node~$v$, not all~$k_v$, but exactly~$n$ neighbors are chosen randomly out of all $k_v$ neighbors of $v$. These $n$ neighbors are scheduled to visit, but only if they have not been visited before.

Respondent-Driven Sampling (RDS)~\cite{Heckathorn97_RDS_introduction,Salganik2004,VolzHeckathorn08} adopts SBS to penetrate hidden populations (such as that of drug addicts) in social surveys.
In Section~\ref{sec:Related_Work}, we comment on current techniques to correct for SBS/RDS bias towards nodes of higher degree.

%
%
%
%
%

\section{Graph model $RG(p_k)$}\label{sec:Graph model}

\begin{table}
  \centering
{\footnotesize
\begin{tabular}{l|l}
  \hline
  $G=(V,E)$ & graph $G$ with nodes $V$ and edges $E$\\
  $k_v$ & degree of node $v$\\
  $p_k \ = \frac{1}{|V|}\sum_{v\in V} 1_{k_v=k}$ & degree distribution in $G$\\
  $\av{k}\ =\ \av{p_k} \ = \sum_k k\, p_k$ & average node degree in $G$ \\
  $q_k$ & expected sampled degree distribution\\
  $\av{q_k} \ = \sum_k k\, q_k $ & expected sampled average node degree\\
  $\est{q}_k$ & sampled degree distribution\\
  $\est{p}_k$ & estimated original degree distribution in $G$\\
  $f$   &  fraction of nodes covered by the sample\\
  \hline
\end{tabular}
}
\caption{Notation Summary.}
\label{Tab:notation}\vspace{-0.8cm}
\end{table}

A basic, yet very important property of every graph is its node degree distribution $p_k$, \ie the fraction of nodes with degree equal to~$k$, for all $k\geq0$.\footnote{As we define $p_k$ as a `fraction', not the `probability', $p_k$ determines the degree sequence in the graph, and vice versa.}
Depending on the network, the degree distribution can vary, ranging from constant-degree (in regular graphs), a distribution concentrated around the average value (\eg in Erd\"os-R\'enyi random graphs or in well-balanced P2P networks), to heavily right-skewed distributions with $k$ covering several decades (as this is the case in WWW, unstructured P2P, Internet at the IP and Autonomous System level, OSNs).
We handle all these cases by assuming that we are given \emph{any} fixed node degree distribution $p_k$.
Other than that, the graph~$G$ is drawn uniformly at random from the set of all graphs 
with degree distribution~$p_k$. We denote this model by~$RG(p_k)$. 

Because $RG(p_k)$ mimics an arbitrary node degree distribution~$p_k$, it can be considered a ``first-order approximation'' of real-life graphs. 
Of course, there are many graph properties other than~$p_k$ that are not captured by~$RG(p_k)$. 
However, we show later that, with respect to the BFS sampling bias, $RG(p_k)$ approximates the real Internet topologies surprisingly well.

We use a classic technique to generate $RG(p_k)$, called the \emph{configuration model}~\cite{Molloy1995}: each node~$v$ is given $k_v$ ``stubs'' or ``edges-to-be''.
Next, all these $\sum_{v\in V} k_v= 2|E|$ stubs are randomly matched in pairs, until all stubs are exhausted (and $|E|$ edges are created).
In Fig.~\ref{fig:stubs_on_interval} (ignore the rectangular interval [0,1] for now), we present four nodes with their stubs (left) and an example of their random matching (right).

\section{Analyzing the Node Degree Bias}\label{sec:Analysis}

In this section, we study the node degree bias observed when the graph exploration techniques of Section~\ref{sec:Algorithms} are run on the random graph~$RG(p_k)$ of Section~\ref{sec:Graph model}. In particular, we are interested in the node degree distribution~$q_k$ expected to be observed in the raw sample. Typically, the observed distribution is different from the original one, $q_k\neq p_k$, with higher average value $\av{q_k}>\av{p_k}$  (\ie average sampled and observed node degree, respectively). 
Below, we derive~$q_k$ as a function of~$p_k$ and, in the case of BFS, of the fraction of sampled nodes~$f$.

\subsection{Random walks  (baseline)}
We begin by summarizing the relevant results known for walks, in particular for RW and MHRW. 
They will serve as a reference point for our main analysis of graph traversals below.

\smallskip
\subsubsection{Random Walk (RW)} \label{Weighted sampling with replacements}
Random walks have been widely studied; see \cite{Lovasz93} for an excellent survey. In any given connected and aperiodic graph, the probability of being at a particular node $v$ converges at equilibrium to the stationary distribution~$\pi^\RW_v \eq \frac{k_v}{2|E|}$.
Therefore, the expected observed degree distribution $q^\RW_k$ is
\begin{equation}\label{eq:q_k_RW}
    q^\RW_k \ =\ \ \sum_v \pi^\RW_v \cdot 1_{\{k_v=k\}} \ = \frac{k}{2|E|}\, p_k\,|V| \ =\ \frac{k\, p_k}{\langle k\rangle},
\end{equation}
where $\langle k\rangle$ is the average node degree in~$G$. \eqn{\ref{eq:q_k_RW}} is essentially similar to calculation in~\cite{Feld1991,Newman01_EgoCentered_Networks,Salganik2004,VolzHeckathorn08}.
As this holds for any fixed (and connected and aperiodic) graph, it is also true for all connected graphs generated by the configuration model.
Consequently, the expected observed average node degree is
\begin{equation}\label{eq:Ek_RW}
    \av{q_k^\RW} \ =\ \sum_k k\, q^\RW_k \ =\  \frac{\sum_k k^2\, p_k}{\langle k\rangle}\ =\ \frac{\langle k^2\rangle}{\langle k\rangle},
\end{equation}
where $\langle k^2\rangle$ is the average squared node degree in $G$. We show this value $\frac{\langle k^2\rangle}{\langle k\rangle}$ in Fig.~\ref{fig:contributions}.


%
%

\smallskip
\subsubsection{Metropolis Hastings Random Walk (MHRW)}
It is easy to show that the transition matrix~$P^\MHRW_{u,w}$ shown in \eqn{\ref{eq:P_u,w}} leads to a uniform stationary distribution~$\pi^\MHRW_v \eq \frac{1}{|V|}$~\cite{mcmc-book}, and consequently:
\begin{eqnarray}
\label{eq:q_k_UNI}  q^\MHRW_k &=& p_k \\
\label{eq:Ek_UNI}  \av{q_k^\MHRW} &=& \sum_k k\, q^\MHRW_k \ =\ \sum_k k\, p_k\ =\ \langle k\rangle.
\end{eqnarray}
In Fig.~\ref{fig:contributions}, we show that MHRW estimates the true mean.

\subsection{Graph traversals (Main Result)}

In both RW and MHRW the nodes can be revisited. So the state of the system at iteration~$i\!+\!1$ depends only on iteration~$i$, which makes it possible to analyze with Markov Chain techniques. In contrast, graph traversals do not allow for node revisits, which introduces crucial dependencies between all the iterations and significantly complicates the analysis.
To handle these dependencies, we adopt an elegant technique recently introduced in~\cite{Kim06_poisson_cloning} (to study the size of the largest connected component) and extended in~\cite{Achlioptas05_On_the_bias_of_traceroute_sampling} (to study the bias of traceroute sampling). However, our work differs in many aspects from both \cite{Kim06_poisson_cloning} and \cite{Achlioptas05_On_the_bias_of_traceroute_sampling}, on which we comment in detail in the related work Section~\ref{sec:Related_Work}.

\begin{figure*}[t]
\psfrag{time}[c][c][0.8]{time $t$ (index)}
\psfrag{t}[c][c][0.8]{current time $t$}
\psfrag{v1}[c][c][0.9]{$v_1$}
\psfrag{v2}[c][c][0.9]{$v_2$}
\psfrag{v3}[c][c][0.9]{$v_3$}
\psfrag{v4}[c][c][0.9]{$v_4$}
\includegraphics[width=1\textwidth]{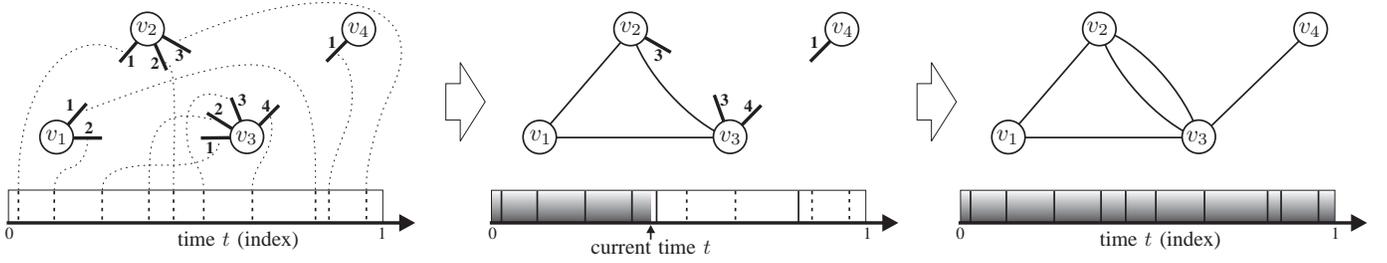}
\caption{An illustration of the stub-level, on-the-fly graph exploration without replacements. In this particular example, we show an execution of BFS starting at node $v_1$.
\quad \textbf{Left:} Initially, each node $v$ has $k_v$ stubs, where $k_v$ is a given target degree of $v$. Each of these stubs is assigned a real-valued number drawn uniformly at random from the interval $[0,1]$ shown below the graph. Next, we follow Algorithm~1 with a starting node $v_1$. The numbers next to the stubs of every node $v$ indicate the order in which these stubs are enqueued on~$Q$.
\quad \textbf{Center:} The state of the system at time $t$. All stubs in $[0,t]$ have already been matched (the indices of matched stubs are set in plain line). All unmatched stubs are distributed uniformly at random on $(t,1]$. This interval can contain also some (here two) already matched stubs.
\quad \textbf{Right:} The final result is a realization of a random graph $G$ with a given node degree sequence (\ie of the configuration model). $G$ may contain self-loops and multiedges.
}
\label{fig:stubs_on_interval}
\vspace{-0.4cm}
\end{figure*}

\smallskip
\subsubsection{Exploration without replacement at the stub level}

We begin by defining Algorithm~1 (below) - a general graph traversal technique that collects a sequence of nodes $S$, without replacements. To be compatible with the configuration model (see Section~\ref{sec:Graph model}), we are interested in the process \emph{at the stub level}, where we consider one stub at a time, rather than one node at a time. An integral part of the algorithm is a queue~$Q$ that keeps the discovered, but still not-yet-followed stubs. First, we enqueue on $Q$ all the stubs of some initial node~$v_1$, and by setting $S\!\gets\![v_1]$. Next, at every iteration, we dequeue one stub from $Q$, call it $a$, and follow it to discover its partner-stub~$b$, and $b$'s owner $v(b)$. If node $v(b)$ is not yet discovered, \ie if $v(b)\notin S$, then we append $v(b)$ to $S$ and we enqueue on $Q$ all other stubs of $v(b)$.
\begin{algorithm}[h!]
\caption{Stub-Level Graph Traversal}
\label{alg:1}
\begin{algorithmic}[1]
\STATE $S\gets [v_1]$ \ and \ $Q\gets$ [all stubs of $v_1$] 
\WHILE {$Q$ is nonempty}
    \STATE Dequeue $a$ from $Q$
    \STATE Discover $a$'s partner $b$
    \IF {$v(b)\notin S$}                       
            \STATE Append $v(b)$ to $S$     
            \STATE Enqueue on $Q$ all stubs of $v(b)$ except $b$
    \ELSE {}
    \STATE Remove $b$ from $Q$
    \ENDIF
\ENDWHILE
\end{algorithmic}
\end{algorithm}

Depending on the scheduling discipline for the elements in~$Q$ (line~3), Algorithm~1 implements BFS (for a first-in first out scheduling), DFS (last-in first-out) or Forest Fire (first-in first-out with randomized stub losses). Line~9 guarantees that the algorithm never tracebacks the edges, \ie that stub $a$ dequeued from $Q$ in line~3 never belongs to an edge that has already been traversed in the opposite direction.

\smallskip
\subsubsection{Discovery on-the-fly}\label{sec:Discovery on the fly}
In line~4 of Algorithm 1, we follow stub $a$ to discover its partner $b$.
In a fixed graph~$G$, this step is deterministic. In the configuration model~$RG(p_k)$, a fixed graph~$G$ is obtained by matching all the stubs uniformly at random. Next, we can sample this fixed graph and average the result over the space of all the random graphs~$RG(p_k)$ that have just been constructed. Unfortunately, this space grows exponentially with the number of nodes~$|V|$, making the problem untractable. Therefore, we adopt an alternative construction of~$G$ - by iteratively selecting~$b$ on-the-fly (\ie every time line~4 is executed), uniformly at random from all still unmatched stubs. By the principle of deferred decisions~\cite{Randomized_Algorithms_book}, these two approaches are equivalent.

With the help of the on-the-fly approach, we are able to write down the equations we need. Indeed, let us denote by $X_i\in V$ the $i$th selected node, and let $\Prob(X_1\eq u)$ be the probability that node $u\in V$ is chosen as a starting node. It is easy to show that with $z\eq 2|E|$ we have
\vspace{-0.2cm}

{\small
\begin{eqnarray}
\label{eq:probX_2} \Prob(X_2\eq v) &=& \sum_{u \neq v} \frac{k_{v}}{z\m k_{u}}\cdot\Prob(X_1\eq u) \\
 \Prob(X_3\eq w) &=& \sum_{v \neq w}\sum_{u \neq w,v} \frac{k_{w}}{z\m k_{v}\m k_{u}} \cdot \frac{k_{v}}{z\m k_{u}}\cdot\Prob(X_1\eq u),\quad
\end{eqnarray}}\vspace{-0.2cm}

\noindent and so on. Theoretically, these equations allow us to calculate the expected node degree at any iteration, and thus the degree bias of BFS.



\smallskip
\subsubsection{Breaking the dependencies}\label{subsec:Breaking the dependencies}
There is still one problem with the equations above. Due to the increasing number of nested sums, the results can be calculated in practice for a first few iterations only.
This is because we select stub~$b$ uniformly and independently at random from all the \emph{unmatched} stubs. So the stub selected at iteration $i$ depends on the stubs selected at iterations $1\ldots i\m 1$, which results in the nested sums.
We remedy this problem by implementing the on-the-fly approach as follows.
First, we assign each stub a real-valued index $t$ drawn uniformly at random from the interval $[0,1]$.
Then, every time we process line~4, we pick $b$ as the unmatched stub with the smallest index. We can interpret this as a continuous-time process, where we determine progressively the partners of stubs dequeued from~$Q$, by scanning the interval from time $t\eq 0$ to $t\eq 1$ in a search of unmatched stubs. Because the indices chosen by the stubs are independent from each other, the above trick breaks the dependence between the stubs, which is crucial for making this approach tractable.
%
%

In Fig.~\ref{fig:stubs_on_interval}, we present an example execution of Algorithm~1, where line~4 is implemented as described above.

\smallskip
\subsubsection{Expected sampled degree distribution $q^\BFS_k$}
Now we are ready to derive the expected observed degree distribution $q_k$. Recall that all the stub indices are chosen independently and uniformly from $[0,1]$.
A vertex~$v$ with degree~$k$ is not sampled yet at time $t$ if the indices of all its $k$ stubs are larger than~$t$, which happens with probability~$(1\m t)^k$.
So the probability that $v$ is sampled before time $t$ is $1\m (1\m t)^k$. Therefore, the expected fraction of vertices of degree $k$ sampled before $t$ is
\begin{equation}\label{eq:f_k(t)}
    f_k(t) = p_k  (1\m (1\m t)^k).
\end{equation}
By normalizing \eqn{\ref{eq:f_k(t)}}, we obtain the expected observed (\ie sampled) degree distribution at time $t$:
\begin{equation}\label{eq:q_k_t}
   q^\BFS_k(t)\ =\  \frac{f_k(t)}{\sum_l f_l(t)}\ =\ \frac{p_k (1 - (1\m t)^k)}{\sum_l p_l (1 - (1\m t)^l)}.
\end{equation}
Unfortunately, it is difficult to interpret $q^\BFS_k(t)$ directly, because~$t$ is proportional neither to the number of matched edges nor to the number of discovered nodes. Recall that our primary goal is to express $q^\BFS_k$ as a function of fraction $f$ of covered nodes.
We achieve this by calculating $f(t)$ - the expected fraction of nodes, of any degree, visited before time $t$
\begin{equation}\label{eq:f(t)}
    f(t) =  \sum_k f_k(t) =  1 - \sum_k p_k (1\m t)^k \ .
\end{equation}
Because $p_k\geq 0$, and $p_k> 0$ for at least one $k>0$, the term $\sum_k p_k (1\m t)^k$ is  continuous and strictly decreasing from 1 to 0 with $t$ growing from 0 to 1. Thus, for $f\in [0,1]$ there exists a well defined $t\eq t(f)$ that satisfies Eq.(\ref{eq:f(t)}), \ie the inverse of $f(t)$.
Although we cannot compute $t(f)$ analytically (except in some special cases such as for $k\leq 4$), it is straightforward to find it numerically. Now, we can rewrite Eq.~(\ref{eq:q_k_t}) as \vspace{-0.1cm}
\begin{equation}\label{eq:q_k_f}
   q^\BFS_k(f)\ =\ \frac{p_k (1 - (1\m t(f))^k)}{\sum_l p_l (1 - (1\m t(f))^l)},
\end{equation}
which is the expected observed degree distribution after covering fraction~$f$ of nodes of graph~$G$. Consequently, the expected observed average degree is
\begin{equation}\label{eq:Mean(q_k)_f}
	\av{q_k^\BFS}(f) \ =\ \sum_k k\cdot q^\BFS_k(f).
\end{equation}
In other words, \eqn{\ref{eq:q_k_f}} and \eqn{\ref{eq:Mean(q_k)_f}} describe the bias of BFS sampling under~$RG(p_k)$, which was our first goal in this paper. Below, we further analyze these equations to get more insights in the nature of BFS bias. 

\smallskip
\subsubsection{Equivalence of traversal techniques under $RW(p_k)$}\label{sec:Equivalence_BFS_DFS}
An interesting observation is that, under the random graph model $RW(p_k)$, all common traversal techniques (BFS, DFS, FF, SBS, etc) are subject to exactly the same bias. The explanation is that the sampled node sequence~$S$ is fully determined by the choice of stub indices on $[0,1]$, independently of the way we manage the elements in~$Q$.


\smallskip
\subsubsection{Equivalence of traversals to weighted sampling without replacement}\label{sec:Equivalence to weighted sampling}
Consider a node $v$ with a degree $k_v$. The probability that $v$ is discovered before time $t$, given that it has not been discovered before~$t_0\leq t$, is \vspace{-0.2cm}
\begin{equation}\label{eq:v_before_t}
    \Prob(\textrm{$v$ before time $t$ $|$ $v$ not before $t_0$}) = 1-\left(\frac{1\m t}{1\m t_0}\right)^{k_v}
\end{equation}
We now take the derivative of the above equation with respect to $t$,
which results in the conditional probability density function $k_v(\frac{1\m t}{1\m t_0})^{k_v\m 1}$.
Setting $t\!\!\rightarrow\! t_0$ (but keeping $t\!>\!\!t_0$), reduces it to $k_v$, which is the probability density that $v$ is sampled at $t_0$, given that it has not been sampled before.
This means that at every point in time, out of all nodes that have not yet been selected, the probability of selecting $v$ is proportional to its degree $k_v$. Therefore, this scheme is equivalent to  node sampling weighted by degree, without replacements. 

\smallskip
\subsubsection{Equivalence of traversals with $f\!\!\rightarrow\! 0$ to RW}\label{sec:Equivalence to RW}
Finally, for $f\!\!\rightarrow\! 0$ (and thus $t\!\!\rightarrow\! 0$), we have $1\m (1\m t)^k \simeq kt$, and Eq.~(\ref{eq:q_k_t}) simplifies to Eq.~(\ref{eq:q_k_RW}). This means that in the beginning of the sampling process, every traversal technique is equivalent to RW, as shown in Fig. 1 for $f\!\!\rightarrow\! 0$.

\smallskip
\subsubsection{$\av{q_k^\BFS}$ is decreasing in $f$} \label{subsec:k^* is decreasing in f}
As in Section~\ref{sec:Discovery on the fly}, let $X_i\in V$ be the $i$th selected node, and let $z\eq 2|E|$. We have shown above that our procedure is equivalent to weighted sampling without replacements, thus we can write $\Prob(X_1\eq u) = \frac{k_u}{z}$. Now, it follows from Eq.~(\ref{eq:probX_2}) that $\Prob(X_2\eq w) = \frac{k_w}{z}\cdot \alpha_w$, where $\alpha_{w} = \sum_{u \neq w} \frac{k_{u}}{z-k_{u}}$.
Because for any two nodes $a$ and  $b$, we have $\alpha_{b}\m \alpha_{a} = z(k_{a}\m k_{b}) / ((z\m k_{a})(z\m k_{b})),$
$\alpha_{w}$ strictly decreases with growing $k_{w}$. As a result, $\Prob(X_2)$ is more concentrated around nodes with smaller degrees than is $\Prob(X_1)$, implying that $\Mean[k_{X_2}] < \Mean[k_{X_1}]$. We can use an analogous argument at every iteration $i\leq |V|$, which allows us to say that $\Mean[k_{X_{i}}] < \Mean[k_{X_{i-1}}]$. In other words, $\av{q_k^\BFS}(f)$ is a decreasing function of $f$.

A practical consequence is that many short traversals are more biased than a long one, with the same total number of samples.

\subsubsection{Comments on the graph connectivity}
%


Note that the configuration model $RG(p_k)$ might result in a graph $G$ that is not connected. In this case, every exploration technique covers only the component~$C$ in which it was initiated; consequently, the process described in Section~\ref{subsec:Breaking the dependencies} stops once~$C$ is covered. 

In practice, it is also possible to efficiently generate a simple and connected random graph with a given degree sequence~\cite{Viger2005}.


\section{Correcting for node degree bias} \label{sec:Correcting for node degree bias}

In the previous section we derived the expected observed degree distribution~$q_k$ as a function of the original degree distribution~$p_k$. 
The distribution~$q_k$ is usually biased towards high-degree nodes, \ie $\av{q_k}\!>\!\av{p_k}$.
Moreover, because many node properties are correlated with the node degree~\cite{Gjoka2010}, their estimates are also potentially biased. 
For example, let~$x(v)$ be an arbitrary function defined on graph nodes~$V$ (\eg node age) and let its mean value
\begin{equation}
  x_\AV = \frac{1}{|V|}\sum_{v \in V} x(v)
\end{equation}
be the value we are trying to estimate.
If $x(v)$ is somehow correlated with node degree~$k_v$, then the straightforward estimator 
$ \est{x}^{\, naive}_\AV= 1/|S|\cdot \sum_{v \in S} x(v)$
is subject to the same bias as is $\av{q_k}$.
In this section, we derive unbiased estimators~$\est{x}_\AV$ of~$x_\AV$. 
We also directly apply~$\est{x}_\AV$ to obtain the estimators $\est{p}_k$ and $\av{\est{p}_k}$ of the original node degree distribution and its mean, respectively.

Let $S\subset V$ be a sequence of vertices that we sampled. Based on $S$, we can estimate $q_k$ as
\begin{eqnarray}
\label{eq:est{q}_k}   \est{q}_k &=&  \frac{\textrm{number of nodes in $S$ with degree $k$}}{|S|}.  
\end{eqnarray}

%
%
%

\smallskip
\subsection{Random walks (baseline)}

\subsubsection{Random Walk (RW)} \label{subsec:Random Walk (RW)}
Under RW, the sampling probability of a node~$v$ is proportional to its degree~$k_v$. 
Because the sampling is done with replacements, we can apply the Hansen-Hurwitz estimator~\cite{HansenHurwitz1943} to obtain the following unbiased estimator~\cite{Feld1991,Newman01_EgoCentered_Networks,Salganik2004,VolzHeckathorn08}
\begin{equation}\label{eq: est{x}^RW_AV}
	\est{x}^{\,\RW}_\AV\ =\ \frac{\sum_{v \in S} x(v)/k_v}{\sum_{v \in S} 1/k_v}.
\end{equation}
For example, if $x(v)\eq 1_{\{k_v=k\}}$ then $\est{x}^{\,\RW}_\AV$ estimates the proportion of nodes with degree equal to~$k$, \ie exactly~$p_k$. In that case, \eqn{\ref{eq: est{x}^RW_AV}} simplifies to 
\begin{equation}\label{eq:est{p}_k_RW}
    \est{p}_k^{\,\RW}\ =\ \frac{\est{q}_k}{k}\ \cdot\ \left(\sum_l \frac{\est{q}_l}{l}\right)^{-1}
\end{equation}
where we used the fact that $\sum_{v\in S} 1_{\{k_v=k\}} = |V|\cdot\est{q}_k$.
From Eq.(\ref{eq:est{p}_k_RW}), we can estimate the average node degree as
\begin{equation}\label{eq:Ek_est_RW}
   \av{\est{p}_k^{\,\RW}}\ =\ \sum_k k\,\est{p}_k^{\,\RW} \ =\ 1 \cdot \left(\sum_l \frac{\est{q}_l}{l}\right)^{-1}=\frac{|S|}{\sum_{v\in S} \frac{1}{k_v}}
\end{equation}

\smallskip
\subsubsection{Metropolis Hastings Random Walk (MHRW)}
Under MHRW, we trivially have
\begin{eqnarray}
\est{x}^{\,\MHRW}_\AV &=& \frac{1}{|S|} \sum_{v \in S} x(v),\\
  \est{p}_k^{\,\MHRW} &=& \est{q}_k, \\ 
\label{eq:est_k_UNI}    \av{\est{p}_k^{\,\MHRW}} &=&  \sum_k k\,\est{p}_k^{\,\MHRW}  \ =\  \sum_k k\,\est{q}_k. 
\end{eqnarray}

\subsection{Graph traversals}
Under BFS and other traversals, the inclusion probability~$\pi^\BFS_v$ 
(\ie the probability of node~$v$ being included in sample~$S$) 
of node $v\in V$ is proportional to 
$$ \pi^\BFS_v\ \ \sim\ \ \frac{q^{\,\BFS}_{k_v}}{p_{k_v}} \ \ \sim\ \ 1 - (1\m t(f))^{k_v},$$
where the second relation originates from \eqn{\ref{eq:q_k_f}}.
Consequently, an application of the Horvitz-Thompson estimator~\cite{HorvitzThompson1952}, 
designed typically for sampling without replacement, leads to
\begin{equation}
 \est{x}^{\,\BFS}_\AV\ =\ \left(\sum_{v \in S}\frac{x(v)}{1 \m (1\m t(f))^{k_v}}\right) \cdot   \left(\sum_{v \in S}\frac{1}{1 \m (1\m t(f))^{k_v}}\right)^{-1}.
\end{equation}
Now, similarly to the analysis of RW (above), we obtain
%
\begin{eqnarray}
\label{eq:est{p}_k_BFS} \est{p}_k^{\,\BFS} & =& \frac{\est{q}_k}{1 - (1\m t(f))^k}\ \cdot\ \left(\sum_l \frac{\est{q}_l}{1 - (1\m t(f))^l}\right)^{-1}\\
\label{eq:est{k}_BFS} \av{\est{p}_k^{\,\BFS}} & = & \sum_k k\,\est{p}_k^{\,\BFS}.
\end{eqnarray}
However, in order to evaluate these expressions, we need to evaluate $t(f)$, that, in turn, requires $p_k$. We can solve this chicken-and-egg problem iteratively, if we know the real fraction $f^\real$ of covered nodes, or equivalently the graph size~$|V|$. First, we evaluate Eq.(\ref{eq:est{p}_k_BFS}) for some values of $t$ and feed the resulting $\est{p}_k$'s into Eq. (\ref{eq:f(t)}) to obtain the corresponding $f$'s. By repeating this process, we can efficiently drive the values of $f$ arbitrarily close to $f^\real$, and thus find the desired $\est{p}_k$.

In summary, for BFS, we showed how to estimate the mean~$x_\AV$ of an arbitrary function~$x(v)$ defined on graph nodes, with the estimator of the original degree distribution~$p_k$ as a special case.  
Note that our approach is feasible, as it requires only the sample~$S$ (with value $x(v)$ and degree $k_v$ for every node $v\in S$) and the fraction~$f$ of sampled nodes. 
In \cite{kurant_networkx_traversals}, we make a~\texttt{python} implementation of all the above estimators publicly available.


\subsection{Alternative approach}
In Section~\ref{sec:Unbiased BFS estimators}, we propose and evaluate a family of alternative correction procedures that are 
 \emph{unbiased for any arbitrary topology}. 
Although seemingly attractive, they are characterized by large variance, which makes them far less effective than our $RG(p_k)$-based correction technique.

\smallskip


\begin{figure*}
\psfrag{fraction f}[c][b][0.8]{$f$\ \ fraction of covered nodes}
\psfrag{degree}[c][t][0.8]{$\av{q_k}$\ \ observed average node degree}
\psfrag{degree k}[c][b][0.8]{$k$\ \ node degree}
\psfrag{prob}[c][t][0.8]{Prob($k$)}
\psfrag{degreeDistr}[c][c][1]{Degree distribution}
\psfrag{zeroAss}[c][c][1]{Average node degree}
\psfrag{k1}[r][c][0.7]{$\langle k\rangle$}
\psfrag{k2}[r][c][0.7]{$\frac{\langle k^2\rangle}{\langle k\rangle}$}
\psfrag{L1}[l][c][0.55]{real, $p_k$}
\psfrag{L2}[l][c][0.55]{expected, $q_k$}
\psfrag{L3}[l][c][0.55]{RW, sampled, $\est{q}_k$}
\psfrag{L4}[l][c][0.55]{RW, estimate, $\est{p}_k$}
\psfrag{L5}[l][c][0.55]{BFS, $f\eq 0.1$, sampled, $\est{q}_k(f)$}
\psfrag{L6}[l][c][0.55]{BFS, $f\eq 0.1$, estimate, $\est{p}_k(f)$}
\psfrag{L7}[l][c][0.55]{BFS, $f\eq 0.3$, sampled, $\est{q}_k(f)$}
\psfrag{L8}[l][c][0.55]{BFS, $f\eq 0.3$, estimate, $\est{p}_k(f)$}
\includegraphics[width=1.0\textwidth]{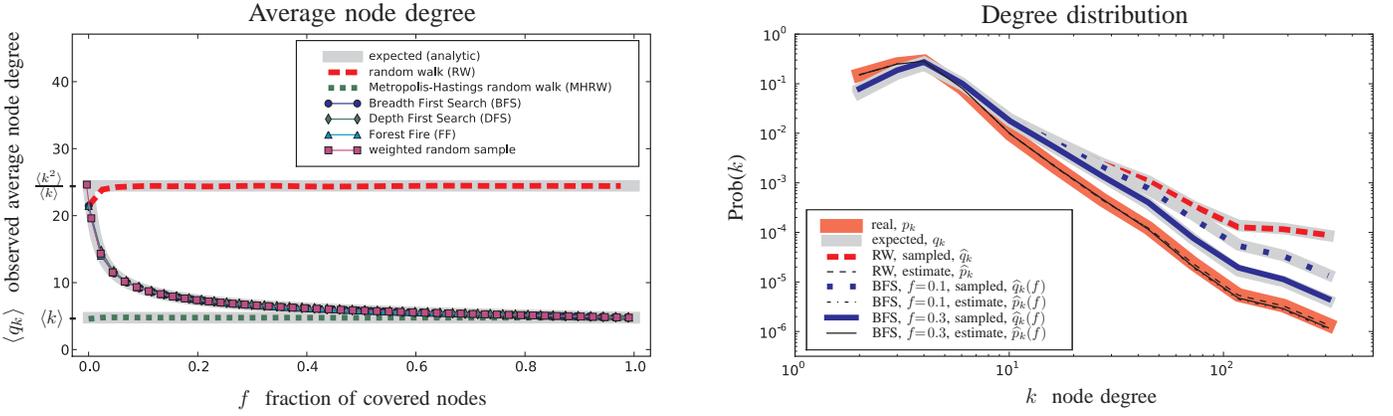}
\caption{{\bf Comparison of sampling techniques in theory and in simulation. }\textbf{Left:}  Observed (sampled) average node degree $\av{q_k}$ as a function of the fraction~$f$ of sampled nodes, for various sampling techniques.
The results are averaged over 1000 graphs with 10000 nodes each, generated by the configuration model with a fixed heavy-tailed  degree distribution~$p_k$ (shown on the right).
\quad \textbf{Right:} Real, expected, and estimated (corrected) degree distributions for selected techniques and values of $f$ (other techniques behave analogously).
\quad We obtained analogous results for other degree distributions and graph sizes $|V|$. The term
$\langle k\rangle$ is the real average node degree, and $\langle k^2\rangle$ is the real average squared node degree.
}
\label{fig:simulations.eps}
\end{figure*}

\begin{figure*}
\psfrag{fraction f}[c][b][0.8]{$f$ - fraction of covered nodes}
\psfrag{degree}[c][t][0.8]{$\av{p_k}$ - average sampled node degree}
\psfrag{degree k}[c][b][0.9]{$k$ - node degree}
\psfrag{P(k)}[c][t][0.9]{$\Prob(k)$}
\psfrag{highAss}[c][c][1]{Average node degree, \ assortativity $r>0$}
\psfrag{lowAss}[c][c][1]{Average node degree, \ assortativity $r<0$}
\psfrag{k1}[r][c][0.7]{$\langle k\rangle$}
\psfrag{k2}[r][c][0.7]{$\frac{\langle k^2\rangle}{\langle k\rangle}$}
\includegraphics[width=1.\textwidth]{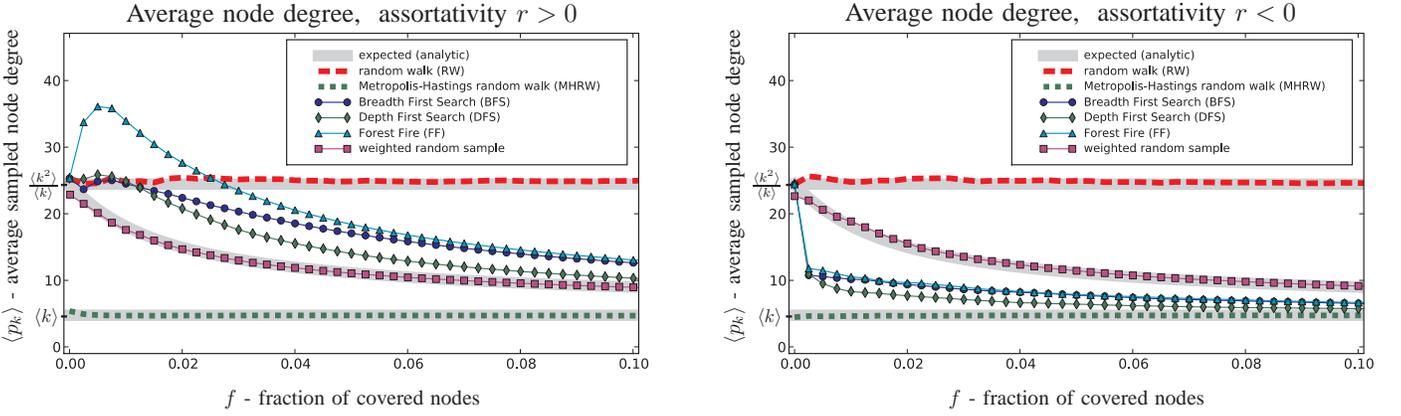}
\caption{{\bf The effect of assortativity $r$ on the results}. First, we use the configuration model with the same degree distribution~$p_k$ as in Fig.~\ref{fig:simulations.eps} (and the same number of nodes $|V|=10000$) to generate a graph $G$. Next, we apply the pairwise edge rewiring technique~\cite{Maslov2002} to change the assortativity $r$ of $G$ without changing node degrees. This technique iteratively takes two random edges $\{v_1,w_1\}$ and $\{v_2,w_2\}$, and rewires them as $\{v_1,w_2\}$ and $\{v_2,w_1\}$ only if it brings us closer to the desired value of assortativity $r$. As a result, we obtain graphs with a positive (left) and negative (right) assortativity $r$. Note that for a better readability, we present only the values of $f\in[0,0.1]$, \ie ten times smaller than in Fig.~\ref{fig:simulations.eps}.
}
\label{fig:assortativity.eps}
\end{figure*}

\section{Simulation results} \label{sec:Simulation}

In this section, we evaluate our theoretical findings on random and real-life graphs. 



\subsection{Random graphs}\label{subsec:Random graphs}

Fig.~\ref{fig:simulations.eps} verifies all the formulae derived in this paper, for the random graph $RG(p_k)$ with a given degree distribution. The analytical expectations are plotted in thick plain lines in the background and the averaged simulation results are plotted in thinner lines lying  on top of them. We observe almost a perfect match between  theory and simulation in estimating the sampled degree distribution $q_k$ (Fig.~\ref{fig:simulations.eps}, right) and its mean~$\av{q_k}$ (Fig.~\ref{fig:simulations.eps}, left). 
Indeed, all traversal techniques follow the same curve (as predicted in Section~\ref{sec:Equivalence_BFS_DFS}), which initially coincides with that of RW (see Section~\ref{sec:Equivalence to RW}) and is monotonically decreasing in~$f$ (see Section~\ref{subsec:k^* is decreasing in f}). 
We also show that degree-weighted node sampling without replacements exhibits exactly the same bias (see Section~\ref{sec:Equivalence to weighted sampling}). 
Finally, applying the estimators $\est{p}_k$ derived in Section~\ref{sec:Correcting for node degree bias} perfectly corrects for the bias of~$q_k$.


\smallskip

Of course, real-life networks are substantially different from~$RG(p_k)$. 
For example, depending on the graph type, 
nodes may tend to connect to similar or different nodes. Indeed, in most social networks high-degree nodes tend to connect to other high-degree nodes~\cite{Newman02}. Such networks are called \emph{assortative}. In contrast, biological and technological networks are typically \emph{disassortative}, \ie they exhibit significantly more high-degree-to-low-degree connections. This observation can be quantified by calculating the \emph{assortativity coefficient}~$r$~\cite{Newman02}, which is the correlation coefficient computed over all edges (\ie degree-degree pairs) in the graph. Values $r\!<\!0$,  $r\!>\!0$ and $r\!=\!0$ indicate disassortative, assortative and purely random graphs, respectively.

For the same initial parameters as in Fig.~\ref{fig:simulations.eps} ($p_k$, $|V|$), we simulated different levels of assortativity.  Fig.~\ref{fig:assortativity.eps} shows the results.
Graph assortativity~$r$ strongly affects the first iterations of traversal techniques. Indeed, for assortativity $r>0$ (Fig.~\ref{fig:assortativity.eps}, left), the degree bias is even stronger than for $r=0$ (Fig.~\ref{fig:simulations.eps}, left). This is because the high-degree nodes are now interconnected more densely than in a purely random graph, and are thus easier to discover by sampling techniques that are inherently biased towards high-degree nodes.
Interestingly, Forest Fire is by far the most affected. A possible explanation is that under Forest Fire, low-degree nodes are likely to be completely skipped by the first sampling wave.
Not surprisingly, a negative assortativity $r<0$ has the opposite effect: every high-degree node tends to connect to low-degree nodes, which significantly slows down the discovery of the former.

In contrast, random walks RW and MHRW are not affected by the changes in assortativity. This is expected, because their stationary distributions hold for \emph{any} fixed (connected and aperiodic) graph regardless of its topological properties.


\subsection{Real-life fully known topologies}\label{subsec:Real-life fully known topologies}

Recall, that our analysis is based on the random graph model~$RG(p_k)$ (see~Section~\ref{sec:Graph model}), which is only an approximation of a typical real-life network~$G$. Indeed, $RG(p_k)$ follows the node degree distribution of $G$, but is likely to miss other important properties such as assortativity~\cite{Newman02}, whose effect on the BFS process we have just demonstrated. 
For this reason, one may expect that the technique based on $RG(p_k)$ performs poorly on real-life graphs. Surprisingly, this is not the case.  

\begin{figure*}
\psfrag{k0}[l][c][0.8]{Average node degree:}
\psfrag{k1}[l][c][0.6]{$\av{p_k}$ - real}
\psfrag{k2}[l][c][0.6]{$\av{q^\BFS_k}$ - expected by BFS}
\psfrag{k3}[l][c][0.6]{$\av{\est{q}^{\,\BFS}_k}$ - sampled by BFS}
\psfrag{k4}[l][c][0.6]{$\av{\est{p}^{\,\BFS}_k}$ - corrected by BFS}
\psfrag{k5}[l][c][0.6]{$\av{\est{p}^{\,\RW}_k}$ - corrected by RW}
\psfrag{frac}[c][b][0.8]{fraction $f$}
\psfrag{degree}[c][c][0.8]{Average degree}
\psfrag{d}[c][c][0.6]{$\langle k\rangle$}
\includegraphics[width=1.\textwidth]{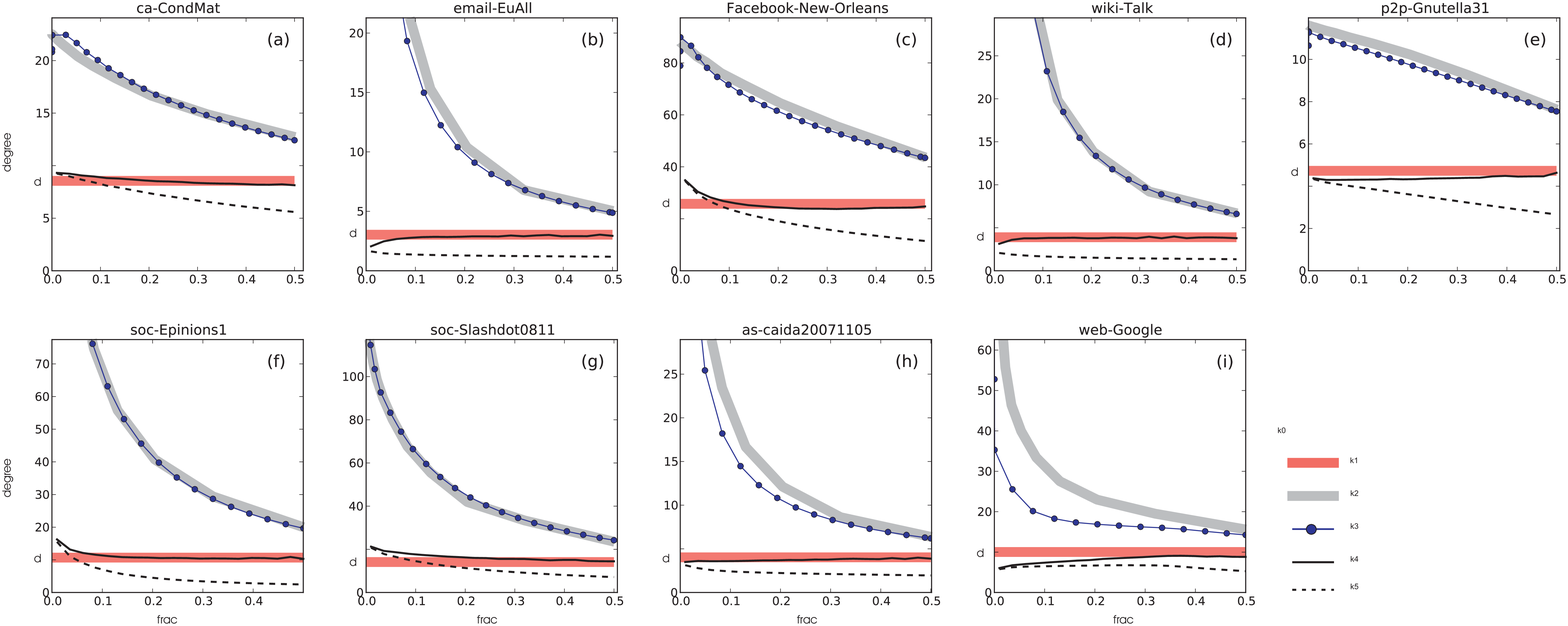}
\caption{{\bf BFS in real-life (fully known) Internet topologies described in Table~\ref{tab:Real-life topologies}.}  
The blue circles represent the average node degree $\av{\est{q}_k^{\,\BFS}}$ sampled by BFS, as the function of the fraction of covered nodes~$f$. 
The thin lines are the corrected values~$\av{\est{p}_k^{\,\BFS}}$ resulting from the BFS  estimator~\eqn{\ref{eq:est{k}_BFS}} (plain line) and the RW estimator \eqn{\ref{eq:Ek_est_RW}} (dashed).
Results are averaged over 1000 randomly seeded BFS samples. 
The thick lines are the analytical expectations assuming the random graph model~$RG(p_k)$. Thick red line (top) is the expectation of $\av{q_k^{\,\BFS}}$, calculated with \eqn{\ref{eq:Mean(q_k)_f}} given the knowledge of the true node degree distribution~$p_k$. 
Thick gray line (bottom) is the expectation of corrected $\av{\est{p}_k^{\,\BFS}}$, \eqn{\ref{eq:est{k}_BFS}}, \ie precisely $\langle k\rangle$.
}
\label{fig:known_topologies}
\end{figure*}

We evaluated our approach on a broad range of large, real-life, fully known Internet topologies. 
As our main source of data we use SNAP Graph Library~\cite{WWW_SNAP_Graph_Library}; Table~\ref{tab:Real-life topologies} overviews these datasets. 
We present the results in Fig.~\ref{fig:known_topologies}. 
Interestingly, in most cases the sampled average node degree~$\av{\est{q}_k^{\,\BFS}}$ closely matches the prediction $\av{q_k^{\,\BFS}}$ of the random graph model~$RG(p_k)$.
%
More importantly, applying our BFS estimator $\av{\est{p}_k^{\,\BFS}}$ of real average node degree
corrects for the bias of $\av{\est{q}_k^{\,\BFS}}$ surprisingly well. Some significant differences are visible only for $f\!\rightarrow\!0$ and for some specific topologies (the last two in Fig.~\ref{fig:known_topologies}), which is exactly because the real-life graphs are not fully captured by graph model~$RG(p_k)$. 

Finally, we also study the RW estimator \eqn{\ref{eq:Ek_est_RW}}, as a simpler alternative to the BFS one~~\eqn{\ref{eq:est{k}_BFS}}. Although they coincide for $f\!\rightarrow\!0$, the RW estimator systematically and significantly underestimates the average node degree $\langle k\rangle$ for larger values of $f$.

\begin{table*}[t!]
  \centering
  {\footnotesize
\begin{tabular}{|r|r|r|r|r|l|}
\hline
    Dataset          & \# nodes   & \# edges & \!$\av{k}\eq\av{p_k}$\!\!  & $\frac{\langle k^2\rangle}{\langle k\rangle}$ & Description \\		
\hline
          ca-CondMat &      21\,363 &      91\,341 &  8.6 &  22.5 & Collaboration network of Arxiv Condensed Matter \cite{Leskovec2007}\\
         email-EuAll &     224\,832 &     340\,794 &  3.0 & 567.9 & Email network of a large European Research Institution \cite{Leskovec2007}\\ 
Facebook-New-Orleans &      63\,392 &     816\,885 & 25.8 &  88.1 & Facebook New Orleans network~\cite{Viswanath2009}\\    
           wiki-Talk &    2\,388\,953 &    4\,656\,681 &  3.9 & 2705.4 & Wikipedia talk (communication) network \cite{Leskovec2010}\\
      p2p-Gnutella31 &      62\,561 &     147\,877 &  4.7 &  11.6 & Gnutella peer to peer network from August 31 2002 \cite{Leskovec2007}\\
       soc-Epinions1 &      75\,877 &     405\,738 & 10.7 & 183.9 & Who-trusts-whom network of Epinions.com \cite{Richardson2003}\\
    soc-Slashdot0811 &      77\,360 &     546\,486 & 14.1 & 129.9 & Slashdot social network from November 2008 \cite{Leskovec2009}\\
    as-caida20071105 &      26\,475 &      53\,380 &  4.0 & 280.2 & CAIDA AS Relationships Datasets, from November 2007 \\
          web-Google &     855\,802 &    4\,291\,351 & 10.0 & 170.4 & Web graph from Google \cite{Leskovec2009}\\                                          
\hline
\end{tabular}}
  \caption{Real-life Internet topologies used in simulations. All graphs are connected and undirected (which required preprocessing in some cases).  }
  \label{tab:Real-life topologies}
\end{table*}

\begin{figure*}
\psfrag{degree k}[c][b][0.8]{$k$\ \ node degree}
\psfrag{prob}[c][t][0.8]{Prob($k$)}
\psfrag{A}[c][t][1]{(a)}
\psfrag{B}[c][t][1]{(b)}
\psfrag{C}[c][t][1]{(c)}
\psfrag{D}[c][t][1]{(d)}
\psfrag{l1}[l][c][0.6]{$p_k$ - real node degree distribution}
\psfrag{l2}[l][c][0.6]{$q^{\,\BFS}_k$ - expected degree distribution}
\psfrag{l3}[l][c][0.6]{$\est{q}^{\,\BFS}_k$ - sampled degree distribution}
\psfrag{l4}[l][c][0.6]{$\est{p}^{\,\BFS}_k$ - corrected degree distribution}
\psfrag{k1}[l][c][0.6]{$\av{p_k}$}
\psfrag{k2}[l][c][0.6]{$\av{q^{\,\BFS}_k}$}
\psfrag{k3}[l][c][0.6]{$\av{\est{q}^{\,\BFS}_k}$}
\psfrag{k4}[l][c][0.6]{$\av{\est{p}^{\,\BFS}_k}$}
\psfrag{t1}[c][t][0.7]{Facebook, BFS${}_{28}$}
\psfrag{t2}[c][t][0.7]{Facebook, BFS${}_{1}$}
\psfrag{t3}[c][t][0.7]{Node degree distributions in Facebook, BFS${}_{1}$}
\psfrag{t4}[c][t][0.7]{Orkut, BFS${}_{1}$}
\psfrag{frac}[c][b][0.8]{fraction $f$}
\psfrag{degree}[c][c][0.8]{Average degree}
\psfrag{d}[c][c][0.6]{$\langle k\rangle$}
\includegraphics[width=1.0\textwidth]{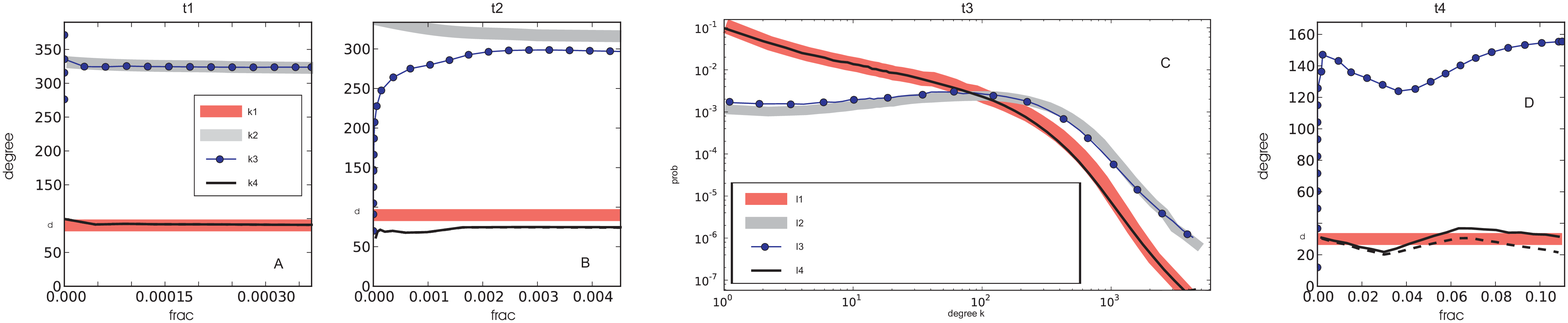}
\caption{{\bf BFS in on-line (not fully known) topologies.}  
As in Fig.~\ref{fig:known_topologies}, except that the plots are based on BFS samples taken in Facebook with 28 (random) seeds~(a) and one seed~(b), as well as in Orkut with one seed~(d). Additionally, we show in~(c) the full node degree distributions for Facebook. 
Because we do not have the true degree distribution~$p_k$ of Orkut, we cannot calculate its analytical curve~$\av{q_k^\BFS}$. Nevertheless, we show in~(d) our best guess of Orkut's average node degree~$\av{k}$ learned by other means, as explained in Footnote~2. 
}
\label{fig:unknown_topologies}
\end{figure*}

\subsection{Sampling Facebook and Orkut}
\label{subsec:Real-life examples: Sampling Facebook and Orkut}

In this section, we apply and test the previous ideas in sampling real-life, large-scale, and not fully known online social networks: Facebook and Orkut. 

\smallskip
\subsubsection{Facebook}

We have implemented a set of crawlers to collect the samples of Facebook (FB) following the BFS, RW, MHRW techniques. 
The data sets are summarized in Table~\ref{tab:Facebook_Datasets}. 
BFS${}_{28}$ consists of 28 small BFS-es initiated at 28 different nodes, which allowed us to easily parallelize the process. Moreover, at the time of data collection, we (naively) thought that this would reduce the BFS bias. After gaining more insight (which, nota bene, motivated this paper), we collected a single large BFS${}_{1}$. 
UNI represents the ground truth.
The details of our implementation are described in~\cite{Gjoka2010,Gjoka2011_Facebook_JSAC}.

\emph{Results.}
We present the Facebook sampling results in Fig.~\ref{fig:unknown_topologies}(a-c) and in  Table~\ref{tab:Facebook_Datasets}.
First, we observe that under BFS${}_{28}$, our estimators $q_k^{\,\BFS}$ and $\est{p}_k^{\,\BFS}$ perform very well. For example, we obtain $\av{\est{p}_k^{\,\BFS}}\eq 85.4$ compared with the true value $\av{k}\eq 94.1$.
In contrast, BFS${}_{1}$ yields $\av{\est{p}_k^{\,\BFS}}\eq 72.7$ only. Most probably, this is because BFS${}_{1}$ consists of a single BFS run that happens to begin in a relatively sparse part of Facebook. Indeed, note that this run starts at $\est{q}_k^{\,\BFS}\eq 50$ for $f\eq 0$, and systematically grows with~$f$ instead of falling. 

Finally, note that both BFS${}_{28}$ and BFS${}_{1}$ are very short compared to the Facebook size, with $f<1\%$ in both cases. For this reason, we observe almost no drop in the sampled average node degre~$\av{q_k^\BFS}$ in Fig.~\ref{fig:unknown_topologies}(a,b). For the same reason, both the BFS and RW estimators yield almost identical results.

All the above observations hold also for the \emph{entire} degree distribution, which is shown in Fig.~\ref{fig:unknown_topologies}(c).

\smallskip
\subsubsection{Orkut}  
\label{subsec:Orkut}
Finally, we apply our methodology to a single BFS sample of Orkut collected in 2006 and described in~\cite{Mislove2007}. It contains $|S|=3072K$ nodes, which accounts for $f\eq 11.3\%$ of entire Orkut size. 

We show the results in Fig.~\ref{fig:unknown_topologies}(d). Similarly to Facebook BFS${}_{1}$, the sampled average node degree $\av{\est{q}_k^{\,\BFS}}$ does not decrease monotonically in $f$. Again, the underlying reason might be the arbitrary choice of the starting node (in sparsely connected India in this case). Nevertheless, the estimator $\av{\est{p}_k^{\,\BFS}}$ approximates the average node degree\footnote{Unfortunately, according to our personal communication with Orkut administrators, there is no ground truth value of the Orkut's average node degree~$\av{k}$ for October 2006, \ie the period when the BFS sample of~\cite{Mislove2007} was collected. 
However, many hints point to a number close to $\av{k}\eq 30$, 
\eg~\cite{Ahn-WWW-07} reports $\av{k}=30.2$ in June-September 2006, and \cite{Anwar2005} reports $\av{k}=19$ in late 2004 (which is in agreement with the densification law~\cite{Leskovec05_Forest_Fire,Leskovec2007}). 
\quad But, as these studies may potentially be subject to various biases, we cannot take these numbers for granted. 
}
 relatively well.

\begin{table}[t!]
  \centering
  {\footnotesize
\begin{tabular}{|c||c|c|c|c|c|}
\hline
Facebook   & UNI              &   RW          &  BFS${}_{28}$   &  BFS${}_{1}$      & MHRW   \\
\hline
$|S|$              &  982K        & 2.26M & 28$\times$81K & 1.19M & 2.26M \\
$f$  &  0.44\%           & 1.03\%         &  28$\times$0.04\%     & 0.54\%  & 1.03\%\\
\hline
  $\av{\est{q}_k}$ & 94.1 & 338.0 & 323.9 & 285.9 & 95.2 \\
  $\av{q_k}$ & - & 329.8  & 329.1  & 328.7  & 94.1  \\
  $\av{\est{p}_k}$ & - & 93.9  &  85.4  & 72.7  & 95.2\\
\hline\hline
Orkut&   &&&&\\
\hline
$|S|$              & -  & - & - & 3.07M & - \\
$f$  &  -           & -  &  -   & 11.3\%  & -\\
\hline
  $\av{\est{p}_k}$ & 30 ${}^2$ &   &  & 33.1 & \\
\hline
\end{tabular}}
  \caption{Facebook and Orkut data sets and measurements.}
  \label{tab:Facebook_Datasets}
\end{table}


%


\section{Arbitrary-topology BFS estimators}\label{sec:Unbiased BFS estimators}

The $RG(p_k)$-based BFS-bias correction procedure is, by construction, unbiased for random graphs~$RG(p_k)$. 
However, when applied to arbitrary graphs, in particular to real-life Internet topologies, our $RG(p_k)$-based estimators are potentially subject to some bias (\ie may be not perfect). 
Fortunately, we have seen in Section~\ref{subsec:Real-life fully known topologies} that this bias is usually very limited. 
This is because~$RG(p_k)$ mimics an arbitrary node degree distribution~$p_k$, which is, by far, the most crucial parameter affecting the BFS degree bias. 





Interestingly, it is possible to derive estimators that are 
\emph{unbiased in any arbitrary topology}.  
Unfortunately, these \emph{arbitrary-topology estimators} are characterized by a very large variance, which makes them, in practice, less effective than the $RG(p_k)$-based estimators. 

In this section we show examples of arbitrary-topology estimators and compare them with $RG(p_k)$-based estimators in simulations.

\subsection{Goal}
Let $G=(V,E)$ be a connected undirected graph. 
A typical (incomplete) graph traversal, such as BFS, is determined by the first node. So we can denote by $S(v)\subset V$ the set of sampled nodes, given that we started at node~$v\in V$. 
Our goal is to use $S(v)$ to estimate the total
\[
  x_\TOT = \sum_{v \in V} x(v) \, ,
\]
where~$x$ is a finite measurable function defined on graph nodes.

\subsection{General arbitrary-topology estimator}
Let $U\in V$ be a random variable representing the first node in our sample, following the probability distribution 
$$\Pr[U\eq w]\ =\ p(w)\ >\ 0.$$ 
Let $Q(w) \subseteq V$ be a set of nodes uniquely defined by $G$ and~$w$.
Define
\begin{equation} \label{eq:est tot}
\est{x}_\TOT = \sum_{v \in Q(U)} \frac{x(v)}{\pi(v)},
\end{equation}
where
\begin{equation} \label{eq:est tot pi}
     \pi(v) = \sum_{w \in V:\ v\in Q(w)} p(w).
\end{equation}

\begin{lemma}
  $\est{x}_\TOT$ is an unbiased estimator of $x_\TOT$.
\end{lemma}
\emph{Proof:}\  In order to prove Lemma~1, we have to show that
 $\Mean[\est{x}_\TOT] = \sum_{v \in V} x(v).$
Indeed:
\begin{align*}
\Mean[\est{x}_\TOT] & = \sum_{w \in V} p(w) \sum_{v \in Q(w)} \frac{x(v)}{\pi(v)} \ =\\
 & =  \sum_{ v \in V}\  \sum_{ w \in V:\ v\in Q(w)}   \frac{x(v)}{\pi(v)} p(w) \ =\\
 & =  \sum_{ v \in V}  \frac{x(v)}{\pi(v)} \sum_{ w \in V:\ v\in Q(w)}   p(w) \ =\\
 & =  \sum_{ v \in V}  \frac{x(v)}{\pi(v)} \pi(v) \ =\\
 & =  \sum_{ v \in V}  x(v).
\end{align*}
(Note that the sums were swapped and appropriately updated after the first step.)
\begin{flushright}
$\boxempty$
\end{flushright}
\bigskip

\subsection{Practical requirements}
We have just shown that $\est{x}_\TOT$ in \eqn{\ref{eq:est tot}} is an unbiased estimator of~$x_\TOT$. 
This is true for \emph{any choice} of $Q(w) \subseteq V$, regardless of our sampling method. By defining~$Q(w)$, we define the estimator. 
However, there are two requirements that we should take into account.

First, our estimator must be \emph{feasible}, \ie we must be able to calculate~$\est{x}_\TOT(v)$ from our sample $S(U)$. This means that all nodes whose values are needed to calculate $\est{x}_\TOT$ must be known (sampled). 
One obvious necessary condition is that $Q(U)\subset S(U)$, because~$Q(U)$ is the set of nodes whose values~$x(v)$ are used in the estimator~$\est{x}_\TOT$ in~\eqn{\ref{eq:est tot}}. 
However, usually we have to know many nodes from beyond $Q(U)$ in order to evaluate~\eqn{\ref{eq:est tot pi}}.
We give some examples below.

Second, the estimator $\est{x}_\TOT$ should be characterized by a \emph{small variance}.

\subsection{Arbitrary-topology estimators for BFS}

Let $B_i(u)$ be a ball of size~$k$ around vertex $u \in V$, \ie the set of all vertices within $i$ hops from $u$. For simplicity, we define our sampling technique as a $i$-stage BFS, \ie $S(u)=B_i(u)$. Depending on our choice of $Q(u)$, we may obtain various feasible arbitrary-topology estimators:

\subsubsection{Trivial} The simplest choice of $Q(v)$ is 
$$Q(v) = \{v\}.$$
This estimator makes use of the first sampled node only, which naturally results in a huge variance.   

\subsubsection{Extreme} We can extend trivial for one specific node $v^*$ to obtain  
$$Q(v) = \left\{ \begin{array}{cl}
                      B_i(v) &\textrm{if\quad $v=v^*$} \\
                      \{v\} &\textrm{otherwise.}
                    \end{array} \right. $$

\subsubsection{Half-radius} A more balanced approach is 
$$Q(v) = B_{\lfloor i/2 \rfloor}(v).$$
In other words, out of the collected $i$-stage BFS sample $S(v)$, we use for estimation only the nodes collected in the first $i/2$ stages of our BFS. It is easy to verify that the half-radius estimator is feasible. 
                    

\subsubsection{Half-radius extended} Finally, we can extend the half-radius estimator to potentially cover some more nodes, as follows.  
$$Q(u) = B_{\lfloor k/2 \rfloor}(u)\ \cup\ \{v\in V:\  B_{i}(v)\subseteq B_i(u) \}.$$


\subsection{Evaluation}

We have tried the above approaches in simulations to estimate the average node degree~$\av{k}=x_\TOT/|V|$.\footnote{For simplicity, we considered the total number of nodes~$|V|$ as known.} 
As our error metric, we used Root Mean Square Error (RMSE), which is appropriate in our case, as it captures both the estimator bias and its variance.
RMSE is defined as:
$$\RMSE\ =\ \sqrt{\Mean\left[(\est{x}_\TOT/|V|-\langle k\rangle)^2\right]}.$$
In our simulations, we calculated the mean $\Mean$ over 1000 BFS samples initiated at nodes chosen uniformly at random, \ie with probability $p(v)=1/|V|$. 
In Table~\ref{tab:unbiased}, we show the results for the half-radius estimator with $i\eq 2$. Other values of~$i$ and other estimators do not improve the results compared to the $RG(p_k)$-based estimator.

Although unbiased, all the proposed arbitrary-topology estimators have very large RMSE compared to the $RG(p_k)$-based estimators. 
There are two main reasons for that. 
First, in order to guarantee feasibility, we usually have $|Q(v)|\ll|S(v)|$, which results in a ``waste'' of values $x(v)$ of most of the sampled nodes. 
Second, the sizes $|Q(v)|$ may significantly differ for different nodes~$v$, which translates to differences in particular estimates~$\est{x}_\TOT(v)$.

%
%
%

To summarize, the arbitrary-topology estimator is unbiased but has a huge variance, which makes it much worse than the potentially slightly biased (for real-life topologies) but much more concentrated $RG(p_k)$-based estimator. 
It is an instance of the well-known ``accuracy vs precision'' trade-off. Indeed, in the statistics terminology, we could say that the arbitrary-topology estimator is ``accurate but very imprecise'', whereas the $RG(p_k)$-based estimator is ``slightly inaccurate but precise''.

\begin{table}[t!]
  \centering
  {\footnotesize
\begin{tabular}{|r|c|c|c|c|}
\hline
   Dataset   & $\av{p_k}$ & correction method         & $\av{\est{p}_k}$  & RMSE  \\		
\hline \multirow{2}{*}{ca-CondMat}	& \multirow{2}{*}{8.6}  & arbitrary-topology	 &8.5	 &10.3	 \\  & & $RG(p_k)$-based & 7.6	 &3.3	 \\  
\hline \multirow{2}{*}{email-EuAll}	& \multirow{2}{*}{3.0}  & arbitrary-topology	 &3.1	 &17.3	 \\  & & $RG(p_k)$-based & 1.7	 &1.5	 \\  
\hline \multirow{2}{*}{Facebook-New-Orleans}	& \multirow{2}{*}{25.8}  & arbitrary-topology	 &25.6	 &33.5	 \\  & & $RG(p_k)$-based & 21.5	 &11.8	 \\  
\hline \multirow{2}{*}{wiki-Talk}	& \multirow{2}{*}{3.9}  & arbitrary-topology	 &3.8	 &27.9	 \\  & & $RG(p_k)$-based & 2.4	 &1.9	 \\  
\hline \multirow{2}{*}{p2p-Gnutella31}	& \multirow{2}{*}{4.7}  & arbitrary-topology	 &4.8	 &4.6	 \\  & & $RG(p_k)$-based & 3.7	 &1.6	 \\  
\hline \multirow{2}{*}{soc-Epinions1}	& \multirow{2}{*}{10.7}  & arbitrary-topology	 &10.3	 &29.3	 \\  & & $RG(p_k)$-based & 9.7	 &6.6	 \\  
\hline \multirow{2}{*}{soc-Slashdot0811}	& \multirow{2}{*}{14.1}  & arbitrary-topology	 &14.5	 &40.5	 \\  & & $RG(p_k)$-based & 17.3	 &6.8	 \\  
\hline \multirow{2}{*}{as-caida20071105}	& \multirow{2}{*}{4.0}  & arbitrary-topology	 &3.9	 &4.7	 \\  & & $RG(p_k)$-based & 2.9	 &1.5	 \\  
\hline \multirow{2}{*}{web-Google}	& \multirow{2}{*}{10.0}  & arbitrary-topology	 &10.6	 &55.2	 \\  & & $RG(p_k)$-based & 6.1	 &5.1	 \\  
\hline
\end{tabular}}
  \caption{Comparison of the arbitrary-topology estimator derived in this section with the $RG(p_k)$-based estimator proposed in the paper.  
  We used the real-life Internet topologies described in Table~\ref{tab:Real-life topologies}. Here, we use the half-radius arbitrary-topology estimator with depth~$i=2$. The results are averaged over 1000 seed nodes chosen uniformly at random from the graph.}
  \label{tab:unbiased}
\end{table}

%
%
%
%
%

\section{Practical recommendations}\label{sec:Recommendations}

In order to sample \emph{node properties}, we recommend using RW. 
RW is simple, unbiased for arbitrary topologies (assuming that we use correction procedures summarized in Section~\ref{subsec:Random Walk (RW)}), and practically unaffected by the starting point. RW is also typically more efficient than MHRW~\cite{Rasti09-RDS,Gjoka2010,Gjoka2011_Facebook_JSAC}.

In contrast, RW and MHRW are not useful when sampling \emph{non-local graph properties}, such as the graph diameter or the average shortest path length. In this case, BFS seems very attractive, because it produces a full view of a particular region in the graph, which is usually a plausible graph for which the non-local properties can be easily calculated. 
However, all such results should be interpreted very carefully, as they may be also strongly affected by the bias of BFS. 
For example, the graph diameter drops significantly with growing average node degree of a network.
Whenever possible, it is a good practice to restrict BFS to some well defined community in the sampled graph. If the community is small enough, we may be able to exhaust it (at least its largest connected component), which automatically makes our BFS sample representative of this community. For example, \cite{Wilson09,Viswanath2009} collected full samples of several Facebook regional networks, and \cite{Albert-Nature-1999,Leskovec2009}~completely covered the WWW graph restricted to one or few domains. 
When such communities are not available (\eg regional networks are not accessible anymore in Facebook), we are left with a regular unconstrained BFS sample. In that case, we recommend applying the $RG(p_k)$-based correction procedure presented in this paper to quantify the node degree bias, which may help us evaluate the bias introduced in the topological metrics.


\section{Conclusion}\label{sec:Conclusion}


To the best of our knowledge, this is  the first work to quantify the node-degree bias of BFS. In particular, we calculated the node degree distribution $q_k$ expected to be observed by BFS as a function of the fraction~$f$ of covered nodes, in a random graph $RG(p_k)$ with a given degree distribution~$p_k$.
We found that for a small sample size, $f\!\rightarrow\!0$, BFS has the same bias as the classic Random Walk, and with increasing~$f$, the bias monotonically decreases.

Based on our theoretical analysis, we proposed a practical $RG(p_k)$-based procedure to correct for this bias when calculating any node statistics. 
Our technique performed very well on a broad range of Internet topologies.
Its ready-to-use implementation can be downloaded from~\cite{kurant_networkx_traversals}.


In this paper, we used our $RG(p_k)$-based correction procedure to estimate local graph properties, such as node statistics. 
An interesting direction for future is to exploit the node degree-biases calculated here to develop estimators of non-local graph properties, such as graph diameter.




\section*{Acknowledgments}
We would like to thank Bruno Ribeiro for useful discussions and the initial idea of the unbiased estimator in Section~\ref{sec:Unbiased BFS estimators};  
Alan Mislove for custom-prepared Orkut BFS sample;
and Minas Gjoka for collecting the Facebook BFS sample. 



\end{document}